\documentclass[aps,prd,preprint,a4paper,showpacs,nofootinbib,superscriptaddress]{revtex4-2}
\usepackage{bm}
\usepackage{indentfirst}
\usepackage{amsmath}
\usepackage{graphicx}
\usepackage{float}
\usepackage{amssymb}
\usepackage{subfigure}
\usepackage{amssymb}
\usepackage{hyperref}
\usepackage{epstopdf}
\usepackage[section]{placeins}

\usepackage[utf8]{inputenc}
\hypersetup{
    colorlinks=true,
    linkcolor=red,
    citecolor=blue,
}
\usepackage{color}
\usepackage[T1]{fontenc}
\usepackage{txfonts}
\usepackage{orcidlink}

\usepackage{adjustbox,lipsum}

\newcommand{\Rmnum}[1]{\expandafter\@slowromancap\romannumeral #1@} 
\newcommand{\bq}{\begin{equation}}
\newcommand{\eq}{\end{equation}}
\newcommand{\bqn}{\begin{eqnarray}}
\newcommand{\eqn}{\end{eqnarray}}
\newcommand{\nb}{\nonumber}

\makeatother

\begin{document}
\title{On Hyperboloidal Foliations in the Study of Black Hole Quasinormal Modes}

\author{Shui-Fa Shen}
\affiliation{School of Intelligent Manufacturing, Zhejiang Guangsha Vocational and Technical University of Construction, 322100, Jinhua, Zhejiang, China}
\affiliation{Postdoctoral Department, Turon University, Karshi 180100, Uzbekistan}

\author{Guan-Ru Li}
\affiliation{Faculdade de Engenharia de Guaratinguet\'a, Universidade Estadual Paulista, 12516-410, Guaratinguet\'a, SP, Brazil}

\author{Xiao-Mei Kuang}
\affiliation{Center for Gravitation and Cosmology, College of Physical Science and Technology, Yangzhou University, Yangzhou 225009, China}

\author{Wei-Liang Qian}\email[E-mail: ]{wlqian@usp.br}
\affiliation{Escola de Engenharia de Lorena, Universidade de S\~ao Paulo, 12602-810, Lorena, SP, Brazil}
\affiliation{Faculdade de Engenharia de Guaratinguet\'a, Universidade Estadual Paulista, 12516-410, Guaratinguet\'a, SP, Brazil}
\affiliation{Center for Gravitation and Cosmology, College of Physical Science and Technology, Yangzhou University, Yangzhou 225009, China}

\author{Ramin G. Daghigh}
\affiliation{Natural Sciences Department, Metropolitan State University, Saint Paul, Minnesota, 55106, USA}

\author{Jodin C. Morey}
\affiliation{Le Moyne College, Syracuse, New York, 13214, USA}

\author{Michael D. Green}
\affiliation{Mathematics and Statistics Department, Metropolitan State University, Saint Paul, Minnesota, 55106, USA}

\author{Rui-Hong Yue}
\affiliation{Center for Gravitation and Cosmology, College of Physical Science and Technology, Yangzhou University, Yangzhou 225009, China}

\begin{abstract}
In this work, we demonstrate that the hyperboloidal foliation technique, applied to the study of black hole quasinormal modes, where the spatial boundary is shifted from spacelike infinity to the future event horizon and null infinity, is effectively equivalent to the continued fraction approach, in which the asymptotic wave function typically diverges at both ends of spatial infinity.
Specifically, a given hyperboloidal slicing, corresponding to a particular choice of coordinates, always uniquely determines a scheme for extracting the asymptotic form of the wave function at the spatial boundary.
Owing to the mathematical equivalence, it follows that the efficiency and precision observed using the hyperboloidal approach should be attributed, not to avoiding the pathological behavior at the spatial boundaries, but primarily to other factors, such as the use of Chebyshev grids.
\end{abstract}

\date{Oct. 1st, 2025}

\maketitle

\section{Introduction}\label{section1}

The hyperboloidal coordinate was first proposed by Zengino\u{g}lu~\cite{agr-qnm-hyperboloidal-02, agr-qnm-hyperboloidal-03} that the divergence of the wave function for black hole quasinormal modes (QNMs) can be attributed to the fact that the boundary of the master equation lies outside of the light cone.
Due to the causality constraints, any localized initial perturbations must never attain the spatial infinity that leads to the aforementioned divergence, and therefore, it is more appropriate to define the boundary of the problem at the future event horizon and future null infinity.
This idea was accomplished by introducing the hyperboloidal coordinates and transforming the boundary of the problem to the null infinity. 
Hyperboloidal coordinates furnish an intuitive geometric picture for the black hole perturbations.

The specific forms of the hyperboloidal coordinates have been further elaborated in~\cite{agr-qnm-hyperboloidal-05, agr-qnm-hyperboloidal-10}, specified to foliations that possess physically relevant interpretations in a Penrose diagram.
In practice, the approach has been adopted to study the QNMs~\cite{agr-qnm-hyperboloidal-04, agr-qnm-hyperboloidal-05, agr-qnm-hyperboloidal-09, agr-qnm-hyperboloidal-10, agr-qnm-lq-matrix-12} as well as the Regge poles~\cite{agr-qnm-Regge-14} of various gravitational systems (see~\cite{agr-hyperboloidal-review-02} for a review).
It has been argued that the choice of hyperboloidal coordinates avoids the pathological behavior at the original spatial boundary.
As a result, it transforms the QNM calculation into a {\it proper} or {\it genuine} eigenvalue problem, which is preferred over the conventional approach.
The significance of the approach resides in its potential application in black hole spectroscopy~\cite{agr-BH-spectroscopy-review-04} and, in particular, spectral instability~\cite{agr-qnm-instability-07, spectral-instability-review-20}.
Although being physically sound and relevant, we show in what follows that the hyperboloidal approach is mathematically equivalent to the approach based on a space-like boundary and asymptotic waveform factorization.

In this short paper, we demonstrate that the hyperboloidal foliation technique is mathematically equivalent to the continued-fraction method.  
In particular, any given hyperboloidal slicing uniquely determines a procedure for extracting the asymptotic behavior of the wave function at spatial infinity.  
After presenting the general arguments in Sec.~\ref{section2}, several explicit examples are discussed in Sec.~\ref{section3}, followed by concluding remarks in Sec.~\ref{section4}.

\section{The equivalence between the two approaches}\label{section2}

The master equation for black hole QNMs typically possesses the form~\cite{agr-qnm-review-02}
\begin{eqnarray}
\frac{\partial^2}{\partial t^2}\Psi(t, r_*)+\left(-\frac{\partial^2}{\partial r_*^2}+V_\mathrm{eff}\right)\Psi(t, r_*)=0 ,
\label{master_eq_ns}
\end{eqnarray}
where the spatial coordinate $r_*$ is known as the tortoise coordinate, which is related to the radial coordinate $r$ by means of the metric function.  
The wavefunction $\Psi$ and the resulting effective potential $V_\mathrm{eff}$ correspond to specific spherical harmonics, while governed by the given background spacetime metric and the specific type of perturbation.
The equation is simplified by using the Fourier transform, namely, 
\bqn
\Psi(t, r_*)=(2\pi)^{-1}\int d\omega e^{-i\omega t} \psi(\omega, r_*) .
\eqn
As a result, the black hole QNM frequencies, $\omega_n$, are determined by solving the eigenvalue problem in the frequency domain:
\begin{equation}
\frac{d^2\psi(\omega, r_*)}{dr_*^2}+[\omega^2-V_\mathrm{eff}]\psi(\omega, r_*) = 0 . \label{master_frequency_domain}
\end{equation}
Since the Fourier transform has been carried out for time slicing with a given $t$, the boundary of the above eigenvalue problem is at spatial infinity $r_* = \pm\infty$.
The stability of the black hole solution requires $\mathrm{Im}\omega_n\le 0$, which leads to the divergence of the outgoing waveforms $\psi(\omega, r_*)$ at both ends of its boundary.
Such a mathematically cumbersome situation can be handled by appropriately factoring out the asymptotic form $\psi_\mathrm{asp}(\omega, r)$ of the wave function, as performed in Leaver's continued fraction method~\cite{agr-qnm-continued-fraction-01}.  
Specifically, we write
\begin{eqnarray}
\psi(\omega, r)=\psi_\mathrm{asp}(\omega, r) \phi(\omega, r),
\label{eq: Leaver-CF}
\end{eqnarray}
where $\phi$ is regular at spatial infinity.

Nonetheless, it was initially pointed out by Zengino\u{g}lu~\cite{agr-qnm-hyperboloidal-02, agr-qnm-hyperboloidal-03} that the divergence of the wave function is irrelevant simply because the boundary of Eq.~\eqref{master_frequency_domain} lies outside of the light cone.
Causality constraints dictate that any localized initial perturbations must never attain spatial infinity, and therefore, the aforementioned divergence is irrelevant from a physical viewpoint.
As a result, it is more appropriate to define the boundary of the problem at the future event horizon and future null infinity.
This idea is accomplished by introducing the transformation to the hyperboloidal coordinates $(\tau, x)$ from $(t, r_*)$.
Specifically, adopting the notation in the literature~\cite{agr-qnm-hyperboloidal-02, agr-qnm-hyperboloidal-03, agr-qnm-instability-07}, it is convenient to first transform the coordinates into dimensionless quantities $(\overline{t}, \overline{x})$ by introducing a length scale $\lambda$. 
\begin{eqnarray}
\overline{t}=\frac{t}{\lambda},~~~~\overline{x}=\frac{r_{*}}{\lambda},~~~~{\hat{V}_\mathrm{eff}=\lambda^2 V_\mathrm{eff}},
\label{dimensionless_quantities}
\end{eqnarray}
where the choice of $\lambda$ is rather arbitrary, primarily aimed at simplifying the resultant master equation. 
Subsequently, the compactified hyperboloidal coordinates $(\tau,x)$ are defined by\footnote{There is a difference in the sign of $H$ in~\cite{agr-qnm-instability-07} compared to the original proposal~\cite{agr-qnm-hyperboloidal-02, agr-qnm-hyperboloidal-03}.}
\begin{eqnarray}
\overline{t}&=&\tau+H(\overline{x}),\nb\\
\overline{x}&=&G(x).
\label{compactified_hyperboloidal_approach}
\end{eqnarray}
where the function $G$ introduces a spatial compactification, while the height function $H$ is defined to guarantee that the boundary (e.g. at $x_\pm=\pm 1$), for a given $\tau$, is the future event horizon or null infinity.
Specifically, it is required that
\begin{eqnarray}
\partial_xH &\le& 1 ,\nb\\
\lim\limits_{x\to x_\pm } \partial_xH &=& \pm 1,
\label{H_relation}
\end{eqnarray}
where
$\partial_xH \equiv \frac{\partial H\left(G(x)\right)}{\partial x}$.

Under coordinates $(\tau,x)$, the master equation Eq.~\eqref{master_eq_ns} possesses the following form:
\begin{eqnarray}
\left[\left(1-\left(\frac{\partial_{x}H}{\partial_{x}G}\right)^2\right)\partial^2_{\tau}-2\frac{\partial_{x}H}{~(\partial_{x}G)^2}\partial_{\tau}\partial_{x}-\frac{1}{\partial_{x}G}\partial_{x}\left(\frac{\partial_{x}H}{\partial_{x}G}\right)\partial_{\tau}-\frac{1}{\partial_{x}G}\partial_{x}\left(\frac{1}{\partial_{x}G}\partial_{x}\right)+\hat{V}_\mathrm{eff}\right]\Psi(\tau, x)=0 .
\label{transformed_master_equation}
\end{eqnarray}
By plugging in the separation of the variables
\begin{eqnarray}
\Psi(\tau,x)=e^{-i\omega \lambda \tau}\phi(x),
\label{wave_function}
\end{eqnarray}
into the master equation, one finds the Fourier transform of Eq.~\eqref{transformed_master_equation} in hyperboloidal time $\tau$
\begin{eqnarray}
\left[a(\omega,x)+b(\omega,x)\partial_{x}+c(\omega,x)\partial_{xx}\right]\phi(x)=0,
\label{simplified_master_equation}
\end{eqnarray}
where
\begin{eqnarray}
a(\omega,x) &=& \lambda\omega\left(\frac{i(\partial_{x}H)(\partial_{xx}G)}{(\partial_{x}G)^3}+\frac{\lambda\omega(\partial_{x}H)^2-i(\partial_{xx}H)}{(\partial_{x}G)^2}-\lambda\omega\right)+\hat{V}_\mathrm{eff},\nb\\
b(\omega,x) &=& \frac{(\partial_{xx}G)-2i\lambda\omega(\partial_{x}G)(\partial_{x}H)}{(\partial_{x}G)^3},\nb\\
c(\omega,x) &=& -\frac{1}{(\partial_{x}G)^2},
\end{eqnarray}
which is an ordinary second-order differential equation in $x$ defined on an interval $[x_-, x_+]$.
Even though the initial perturbations can never exceed the null infinity, it is worth noting that the boundary condition of Eq.~\eqref{simplified_master_equation} does not correspond to that of a bound state. 
To be more specific, the wave function $\phi(x)$ is regular, but it does not vanish at its boundary $x=x_\pm$.
This is because the choice of hyperboloidal coordinates does not turn the original problem into a bound state eigenvalue problem, indicating that the underlying physics remains that of an open system, characterized by the outgoing wave boundary condition.
In other words, the wave function $\phi(x)$ does not vanish at the new boundary; it is just regular and typically attains finite values.
In what follows, we first elaborate on its equivalence and then provide a few explicit examples.

Firstly, it has been pointed out that the quasinormal frequencies obtained from Eq.~\eqref{simplified_master_equation} are identical to those derived from Eq.~\eqref{master_frequency_domain}.
Secondly, the process of separation of variables carried out in hyperboloidal coordinates ($\tau, x, \lambda$), as given in Eq.~\eqref{wave_function}, can be effectively mapped to its counterpart in the original coordinates ($t, r(r_*)$).  
More specifically, substituting the first line of Eq.~\eqref{compactified_hyperboloidal_approach} into Eq.~\eqref{wave_function}, we find
\begin{eqnarray}
\Psi(\tau,x)=e^{-i\omega t}e^{i\omega \lambda H(\overline{x})}\phi(x).
\label{eq: waveformHyper}
\end{eqnarray}
Observe that the separation of variables not only extracts the temporal dependence (in $t$) but also carries an additional factor of the waveform which is purely spatially dependent. 
Comparing Eqs.\ \eqref{eq: Leaver-CF} and \eqref{eq: waveformHyper} we can take
\begin{eqnarray}
\psi_\mathrm{asp} \equiv e^{i\omega\lambda H(\overline{x})} ,\label{assWF}
\end{eqnarray}
which, due to Eq.~\eqref{H_relation}, simply provides the desirable properties of the asymptotic wave function at the boundary.
Therefore, if one chooses the asymptotic wave function to have the form Eq.~\eqref{assWF}, the master Eq.~\eqref{simplified_master_equation} must be equivalent to Eq.~\eqref{master_frequency_domain}.

This is a good place to mention that Eq.~\eqref{wave_function} first appears in Eq.~(6) of~\cite{agr-qnm-hyperboloidal-03}, where the author mentioned that, in the analysis by Dolan and Ottewill~\cite{agr-qnm-geometric-optics-03, agr-qnm-geometric-optics-04}, a similar form was introduced in the calculation of the QNMs and Regge poles in the Schwarzschild and Kerr spacetime.
The motivation stemmed from the feasibility of expanding the frequency (or multipole number) and wavefunction in inverse powers of $\ell$ (or $\omega$), for a {\it particular} choice of wavefunction rescaling that satisfies Eq.~\eqref{H_relation}.  This was crucial for a specific algorithm for calculating QNMs or Regge poles in the eikonal limit.
In particular, unlike Eq.~\eqref{wave_function}, the exponential of the prefactor given in Eq.~(33) of~\cite{agr-qnm-geometric-optics-04} in the Kerr spacetime is not linearly proportional to $\omega$, which is inevitable as discussed below Eq.~\eqref{factorKerr}.
This paper aims to further elaborate on the following aspects.
Firstly, while the asymptotic form of the wave function is identical at its boundary, either by Leaver's strategy or by redefining the time coordinate, the fact that the two approaches are entirely mathematically equivalent in a general context, rather than a particular scenario, may not be entirely apparent to some readers.
Therefore, it is worthwhile to show the equivalence of the two approaches in terms of the height function $H(\overline{x})$ and the asymptotic wavefunction explicitly.
Secondly and more importantly, the choice of hyperboloidal coordinates does not change the physical nature of the problem.
As elaborated below, it remains an open system and does not turn into a bound state problem. 

\section{A few explicit examples}\label{section3}

In this section, we discuss a few examples that embrace different black hole spacetimes.
Our derivations are two-fold.
On the one hand, for given hyperboloidal coordinates, we will explicitly derive the asymptotic form $\psi_\mathrm{asp}(r)$ that is to be extracted from the wave function defined in the ordinary radial coordinate, as part of the procedure in the continued fraction~\cite{agr-qnm-continued-fraction-01, agr-qnm-continued-fraction-04} or matrix method~\cite{agr-qnm-lq-matrix-02}.
On the other hand, for a given recipe of asymptotic waveform, we give the height and compactified functions, Eqs.~\eqref{dimensionless_quantities}-\eqref{H_relation}, that define the corresponding hyperboloidal foliation.

Let us first consider a Schwarzschild black hole.  The Regge-Wheeler potential is
\bqn
V_\text{eff} = \frac{1 - \sigma}{r^2}\left(\ell (\ell + 1) + \sigma (1 - s^2)\right)
\label{eq:schwpot}
\eqn
The height function and compactification in Eq.~\eqref{compactified_hyperboloidal_approach} are given in
\cite{agr-qnm-instability-07} as
\begin{eqnarray}
H(\sigma) &=& -\frac12\left(\ln \sigma+\ln(1-\sigma)-\frac{1}{\sigma}\right) , \nb\\
G(\sigma) &=& \frac12\left(\frac{1}{\sigma}+\ln(1-\sigma)-\ln\sigma\right) ,
\end{eqnarray}
with $\lambda=4M=2$, $\sigma=2M/r=1/r$.
The boundary is defined at $\sigma=1$ for the horizon and $\sigma=0$ for null infinity.
Plugging this into Eq.~\eqref{assWF}, we have
\begin{eqnarray}
\psi_\mathrm{asp}=e^{i\omega \lambda H(\sigma)} = e^{-i\omega\left(\ln\sigma + \ln(1-\sigma) - \frac{1}{\sigma}\right)} = r^{2i\omega}(r-1)^{-i\omega}e^{i\omega r} .
\end{eqnarray}
This is to be compared to the asymptotic wave function proposed in Leaver's paper~\cite{agr-qnm-continued-fraction-01} (cf. Eq.~(5) and \(\rho = -i\omega\))\footnote{This comparison was made by Macedo and Zengino\u{g}lu in the blog hyperboliod.al/post/minimal-guage/},
\begin{eqnarray}
\psi_\mathrm{asp}(r)= r^{2i\omega}(r-1)^{-i\omega}e^{i\omega (r-1)} .
\end{eqnarray}
It is observed that they differ only by a factor $e^{i\omega}$, which depends on the frequency but not on the spatial coordinates. 
Therefore, the remaining part of the wave function must satisfy the same equation.

This is indeed the case.  Substituting Eq.~\eqref{eq:schwpot} into the master equation Eq.~\eqref{simplified_master_equation}, one obtains
\begin{eqnarray}
&&(x-1)^2x \phi''(x)+(1+x^2(3-4i\omega)+x(-4+8i\omega)-2i\omega)\phi'(x)\nb\\
&&-(1+\ell+\ell^2+s^2(x-1)-x-4i\omega+4ix\omega-8\omega^2+4x\omega^2)\phi(x) = 0 ,
\end{eqnarray}
where $x\equiv 1-\sigma$.
Substituting the Taylor expansion of the {\it spatial} part of the wave function in the hyperboloidal coordinate  $\phi(x)=\sum_{n=0}^\infty a_n x^n$, one finds
\begin{eqnarray}
\alpha_0 a_1+\beta_0 a_0 &=& 0,\nb\\
\alpha_n a_{n+1}+\beta_n a_n +\gamma_n a_{n-1} &=& 0, \ \ \ n=1, 2, \cdots
\end{eqnarray}
where
\begin{eqnarray}
\alpha_n &=& (1+n)(1+n-2i\omega) ,\nb\\
\beta_n &=& s^2-1-\ell-\ell^2-2n^2+2n(4i\omega-1)+4i\omega+8\omega^2 ,\nb\\
\gamma_n &=& n^2-s^2-4i\omega-4\omega^2 ,
\end{eqnarray}
which is essentially the three-term recurrence relation employed in the continued fraction method by Leaver (cf. Eqs.~(6-8) of~\cite{agr-qnm-continued-fraction-01}).
It should be noted that the improved convergence demonstrated in Fig.~7 of~\cite{agr-qnm-hyperboloidal-04}, where the authors compare the convergence of their method to Leaver's continued fraction method, is not attributed to the use of hyperboloidal coordinates, but to a refined analysis of the asymptotic behavior of the expansion coefficients.

The second example is the Reissner-Nordström black hole.
Two specific choices of hyperboloidal coordinates have been given in~\cite{agr-qnm-hyperboloidal-05}, referred to as areal radius fixing and Cauchy horizon fixing gauges. 
Although neither of them is identical to an earlier recipe proposed by Leaver in the radial coordinate~\cite{agr-qnm-continued-fraction-04}, one can derive their counterparts' spatial wavefunction $\psi_\mathrm{asp}(r)$ that can be used to extract the asymptotic behavior at both ends of the boundary, and therefore match perfectly with the conventional approach.
Conversely, we will also show that it is possible to derive the height function $H(x)$ starting from Leaver's continued fraction approach, which gives rise to a valid choice of hyperboloidal foliation.

Let us define a few quantities regarding a Reissner-Nordström black hole of mass $M=1/2$ and charge\footnote{Some redundancy appears in these definitions due to merging notations from different references~\cite{agr-qnm-hyperboloidal-05, agr-qnm-continued-fraction-04}.} $Q$
\begin{eqnarray}
\Delta &=& r^2-r+Q^2 = (r-r_+)(r-r_-),  \nb\\
r_\pm &=& \frac12\left(1\pm\sqrt{1-4Q^2}\right) ,\nb\\
\kappa &=& \frac{r_-}{r_+} ,\nb\\
r_* &=& \int dr\frac{r^2}{\Delta} = r+\frac{r_+^2}{r_+-r_-}\ln(r-r_+)-\frac{r_-^2}{r_+-r_-}\ln(r-r_-) .
\end{eqnarray}

For the areal radius fixing gauge of~\cite{agr-qnm-hyperboloidal-05}, we obtain \footnote{This was derived from Eqs. (7), (27) and (135) of \cite{agr-qnm-hyperboloidal-05}; Note that the height function $h(\sigma)$ in \cite{agr-qnm-hyperboloidal-05} is related to our height function according to $\lambda H(\sigma) = -r_* - \lambda h(\sigma)$. }  
\begin{eqnarray}
\lambda H(\sigma) 
&=& -r-\frac{r_+^2}{r_+-r_-}\ln(r-r_+)+\frac{r_-^2}{r_+-r_-}\ln(r-r_-) + \lambda\left(\frac{1}{\sigma}-(1+\kappa)\ln\sigma\right)\nb\\
&=& r-\frac{r_+^2}{r_+-r_-}\ln(r-r_+)+\frac{r_-^2}{r_+-r_-}\ln(r-r_-)+2\ln r-2\ln r_+, \label{gaugeRF}
\end{eqnarray}
where $\lambda=2r_+$, and $\sigma=\lambda/2r$.
Substituting Eq.~\eqref{gaugeRF} into Eq.~\eqref{assWF}, it is readily shown that the spatial part $\psi_\mathrm{asp}(r)$ has the following asymptotic behavior
\begin{eqnarray}
\lim\limits_{r\to +\infty} \psi_\mathrm{asp}(r) &=& c_{\infty} e^{i\omega (r+\ln r)}, \nb\\
\lim\limits_{r\to r_+} \psi_\mathrm{asp}(r) &=& c_+(r-r_+)^{-i\omega\frac{r_+^2}{r_+-r_-}} ,\label{assRN}
\end{eqnarray}
where the constants
\begin{eqnarray}
c_{\infty} &=& r_+^{-2i\omega},\nb\\
c_+ &=& e^{i\omega r_+}(r_+-r_-)^{i\omega\frac{r_-^2}{r_+-r_-}} .
\end{eqnarray}
It is apparent that besides the coefficients $c_{\infty}$ and $c_+$, the asymptotic behavior of the wavefunction $\psi_\mathrm{asp}(r)$ given on the r.h.s of Eqs.~\eqref{assRN} is precisely identical to those derived in~\cite{agr-qnm-continued-fraction-04} (cf. the expression above Eq.~(4) and Eq.~(6)).

Although the expressions are much longer, similar results are obtained for the Cauchy horizon fixing gauge (Eqs.~(137-139) of~\cite{agr-qnm-hyperboloidal-05}).
Apart from the difference in values of the coefficients and compactified spatial variable ($\sigma$ is no longer fixed at the event horizon but at the Cauchy horizon), the exact asymptotic behavior is obtained, demonstrating the equivalence.

Conversely, starting from Leaver's choice of asymptotic spatial waveform (cf. Eq.~(3) of~\cite{agr-qnm-continued-fraction-04}), 
\begin{eqnarray}
\psi_\mathrm{asp}(r) = r_+ e^{-2i\omega r_+}(r_+-r_-)^{-2i\omega-1} r^{-1}(r-r_-)^{1+i\omega}e^{i\omega r}\left(\frac{r-r_-}{r-r_+}\right)^{-i\omega\frac{r_+^2}{r_+-r_-}} ,
\end{eqnarray}
one can define the following functions
\begin{eqnarray}
H(r) &=& r + \ln (r-r_-) - {\frac{r_+^2}{r_+-r_-}}\ln\left(\frac{r-r_-}{r-r_+}\right),\nb\\
G(x) &=& \frac{r_+ - xr_-}{1-x} ,
\end{eqnarray}
where one takes the trivial scaling $\lambda=1$, and the compactified variable $x$ attains $x=0$ at the event horizon and $x=1$ at spatial infinity, similar to the variable $\sigma$ in the areal radius fixing gauge above.
One notes that Leaver's original choice does not directly furnish a form given by the exponential on the r.h.s. of Eq.~\eqref{assWF}, which dictates that the exponent must be linearly proportional to $\omega$.
In fact, when compared against Eq.~\eqref{wave_function}, we have
\begin{eqnarray}
\Psi(\tau, x)=e^{-i\omega \lambda \tau}f(G(x))\phi(x) ,\label{wave_function_Leaver}
\end{eqnarray}
where 
\begin{eqnarray}
f(r) = r_+ e^{-2i\omega r_+}(r_+-r_-)^{-2i\omega-1} \left(\frac{r-r_-}{r}\right) ,\label{f_prefactor_Leaver}
\end{eqnarray}
only depends on the radial coordinate and is manifestly regular at both ends of the boundary.
The equivalence is only achieved by generalizing Eq.~\eqref{wave_function} to Eq.~\eqref{wave_function_Leaver} by including an additional regular rescaling of the wavefunction via the factor $f(r)$.
We note that the presence of this regular function does not affect the viability of the approach.
Its purpose is to simplify the recurrence relation, in this case giving just four terms.

It was pointed out~\cite{agr-qnm-hyperboloidal-05} that hyperboloidal coordinates provide valuable {\it geometric} insights, where the geometry is imbedded in the conformal areal radius of the compactified spacetime $\rho$, into Leaver's recipe.
As a matter of fact, it can be shown that with the right gauge fixing in hyperboloidal coordinates, the conformal prefactor $\rho^n$ defined in Eq.~(55) of~\cite{agr-qnm-hyperboloidal-05} becomes identical to the function $f(r)$ given in Eq.~\eqref{wave_function_Leaver}.
To be more specific, the recurrence relations Eqs.~(153) derived in~\cite{agr-qnm-hyperboloidal-05} for the areal radius fixing gauge are not equivalent to Leaver's results, since the prefactor is a constant in this case.
On the other hand, the recurrence relations  for the Cauchy horizon fixing gauge in Eqs.~(161) of~\cite{agr-qnm-hyperboloidal-05} are entirely equivalent to Leaver's Eqs.~(9) in~\cite{agr-qnm-continued-fraction-04}\footnote{We thank Rodrigo Macedo for pointing this out to us.}.
In the latter case, the prefactor $\rho^n = \frac{2r_+}{r_+-r_-}\left(\frac{r-r_-}{r}\right)$ in Eq.~(149) of \cite{agr-qnm-hyperboloidal-05} is proportional to $f(r)$ given in Eq.~\eqref{f_prefactor_Leaver}.

Lastly, we consider a Kerr black hole with mass $2M=1$ and angular momentum per unit mass $a$:
\begin{eqnarray}
\Delta &=& r^2-r+a^2 = (r-r_+)(r-r_-),  \nb\\
r_\pm &=& \frac12\left(1\pm\sqrt{1-4a^2}\right) .
\end{eqnarray}

Starting from Leaver's recipe~\cite{agr-qnm-continued-fraction-01} (cf. Eq.~(23)), for the radial sector of the master equation, one can show that it corresponds to the following functions (up to a constant) for the hyperboloidal slicing
\begin{eqnarray}
H(r) &=&  r+\ln(r-r_-) + \frac{r_+}{r_+-r_-}\ln\left(\frac{r-r_-}{r-r_+}\right),\nb\\
G(u) &=&  \frac{r_+ - ur_-}{1-u} ,
\end{eqnarray}
where, again, one chooses $\lambda=1$. 
The compactified variable $u$ attains $u=0$ at the event horizon and $u=1$ at spatial infinity.
The detailed derivation involves some subtleties.
Owing to the infeasibility of separation of variables in the coordinate basis associated with the Regge-Wheeler-Zerilli approach, the derivation of the master equation employed the Newman-Penrose formalism, a powerful and more general framework that utilizes the null tetrad basis to analyze perturbative dynamics.
A key feature of this basis is its alignment with the two repeated principal null directions characterizing the background Petrov type-D spacetime.
As a consequence, the Newman-Penrose quantities typically diverge at the horizon owing to the blow-up of tetrad vectors as $\Delta \to 0$.
This issue can be mitigated by effectively transforming to the Hawking-Hartle tetrad, which corresponds to rescaling the wavefunction $\psi \to \Delta^{-s}\psi$, where $s$ is the spin-weight of the perturbation, together with a conformal transformation of the metric $\Omega^1 \propto 1-u$.
It is worth noting that Teukolsky's derivation applies directly to any type-D background metric, including the Schwarzschild black hole.
In particular, the latter yields the Bardeen-Press equations, which can be shown to be equivalent to the Regge-Wheeler-Zerilli equations after similar transformations.
However, since separation of variables is viable in the Schwarzschild case within the metric formalism, this point is often not addressed in the literature, nor in works involving hyperboloidal coordinates.
Nonetheless, following the line of reasoning above, the hyperboloidal slicing can be implemented only after factoring out from the wavefunction a term (cf. Eq.~(35) of~\cite{agr-qnm-hyperboloidal-10}, Eq.~(23) of~\cite{agr-qnm-continued-fraction-01}):
\begin{eqnarray}
g(G(u)) = \frac{1-u}{r_+ - r_-}\,\Delta^{-s} = (r-r_+)^{-s}(r-r_-)^{-s-1} ,\label{factorKerr}
\end{eqnarray}
which, as discussed above, is not regular at the spatial boundary.
Mathematically, unlike $f(r)$, the function $g(r)$ nontrivially impacts the boundary, and this procedure is indispensable for the hyperboloidal foliation to work out, as the remaining asymptotic wave function will be linearly proportional to $\omega$ in the exponential.

Conversely, deriving the asymptotic wavefunction from a given hyperboloidal coordinate is somewhat tedious but straightforward, and can be carried out analogously.
We present an explicit example for the radial-fixing minimal gauge discussed in~\cite{agr-qnm-hyperboloidal-10}.
The asymptotic wavefunction follows by substituting $\lambda = r_+$, $\sigma=\lambda/r$, $\rho(\sigma) = 1$, and $\mu = 1/2\lambda$ (cf. Eqs.~(14), (26), (61), (63-64) of~\cite{agr-qnm-hyperboloidal-10}) into the height function.
By stripping off the irregular part (cf. Eqs.~(10), (27--28) of~\cite{agr-qnm-hyperboloidal-10}), one obtains
\begin{eqnarray}
\psi_\mathrm{asp}(r)
= e^{-i\omega (r_* + \lambda h_0(\sigma))}
= e^{i\omega r}\left(\frac{r}{r_+}-1\right)^{-i\omega \left(\frac{r_+^2}{\sqrt{1-4a^2}}\right)}
\left(\frac{r}{r_+}-\frac{r_-^2}{r_+^2}\right)^{i\omega \left(\frac{r_-^2}{\sqrt{1-4a^2}}\right)}
\left(\frac{r_+}{r}\right)^{-2i\omega r_+\left(1+\frac{r_-^2}{r_+^2}\right)} .
\end{eqnarray}
It is noted that the coordinate penetrates the event horizon in~\cite{agr-qnm-hyperboloidal-05, agr-qnm-hyperboloidal-10}.
Nonetheless, while evaluating the QNMs, the boundary condition remains at the event horizon, and moreover, the physical interpretation as the causal past of future null infinity, based on which the coordinate is established, becomes irrelevant beyond this point.

\section{Concluding remarks}\label{section4}

Without any doubt, the hyperboloidal foliation furnishes a physically pertinent alternative interpretation for the black hole QNM.
In this note, we elaborate on the equivalence between hyperboloidal foliation for black hole QNMs and the more conventional approach. 
Specifically, given the recipe for hyperboloidal foliation, it is shown that it can always be matched to an approach where an asymptotic waveform is factored out to eliminate divergence at the spatial boundary, which leads to the resulting master equation. 
On the other hand, the converse process does not always seem feasible, as it is subject to the viability of some manipulation so that the frequency in the resulting asymptotic waveform is linear in the exponent. 
However, with a small modification [the function $f$ in Eq.~\eqref{wave_function_Leaver}], Leaver's approach can be written in terms of a hyperboloidal foliation in the cases of the Schwarzschild, Reissner-Nordström, and Kerr black holes.
\begin{enumerate}
    \item From a mathematical perspective, it is equivalent to conventional approaches and therefore does not, by itself, guarantee better efficiency or accuracy.
    \item From a physical perspective, the resulting eigenvalue problem is {\it not} a bound state problem: the resulting wave functions $\phi(x)$ [as solutions of Eq.~\eqref{simplified_master_equation}] are regular but do not vanish at its new boundary, inheriting the dissipative nature of an open system manifested in terms of the outgoing wave boundary condition.
\end{enumerate}
In this context, by itself, the hyperboloidal foliation technique cannot be viewed as mathematically distinct from Leaver's method for the evaluation of QNMs (in contrast to, for example, the matrix method that furnishes a distinct numerical approach).

\section*{Acknowledgements}

We acknowledge insightful discussions with An{\i}l Zengino\u{g}lu, Rodrigo P. Macedo, and Alex V. Viñuales.
We gratefully acknowledge the financial support from Brazilian agencies 
Funda\c{c}\~ao de Amparo \`a Pesquisa do Estado de S\~ao Paulo (FAPESP), 
Funda\c{c}\~ao de Amparo \`a Pesquisa do Estado do Rio de Janeiro (FAPERJ), 
Conselho Nacional de Desenvolvimento Cient\'{\i}fico e Tecnol\'ogico (CNPq), 
and Coordena\c{c}\~ao de Aperfei\c{c}oamento de Pessoal de N\'ivel Superior (CAPES).
This work is supported by the National Natural Science Foundation of China (NSFC).
A part of this work was developed under the project Institutos Nacionais de Ci\^{e}ncias e Tecnologia - Física Nuclear e Aplica\c{c}\~{o}es (INCT/FNA) Proc. No. 464898/2014-5.
This research is also supported by the Center for Scientific Computing (NCC/GridUNESP) of São Paulo State University (UNESP).

\bibliographystyle{h-physrev}
\bibliography{references_qian}

\end{document}